# The effect of quenching from different temperatures on $Bi_{0.88}Sb_{0.12}$ alloy

K. Malik,[1] Diptasikha Das,[1] S. K. Neogi,[1,§] A. K. Deb,[2] Arup Dasgupta[3], S. Bandyopadhyay[1,4]

and

Aritra Banerjee[1,4,a)]

[1]Department of Physics, University of Calcutta, 92 A P C Road, Kolkata 700 009, India

[2]Department of Physics, Raiganj College (University College), Uttar Dinajpur 733 134, India

[3]Physical Metallurgy Group, Indira Gandhi Centre for Atomic Research, Kalpakkam 603102, India

[4]Center for Research in Nanoscience and Nanotechnology, University of Calcutta, JD-2, Sector-III, Saltlake City, Kolkata 700 098, India

[§]Presently at: Centre for Advanced Materials, Indian Association for the Cultivation of Science, Jadavpur, Kolkata-700032, India

## ABSTRACT

Structural, thermal, resistive and magnetic properties of melt quenched $Bi_{0.88}Sb_{0.12}$ alloys are reported. The samples are heated at three different temperatures, followed by rapid quenching in liquid nitrogen. Large temperature difference between liquidus and solidus lines, led to microscopic in-homogeneity in the alloy. The effect of quenching from different temperatures in polycrystalline $Bi_{0.88}Sb_{0.12}$ alloy has been studied. The parameters such as strain, unit cell volume, and resistivity are found to increase with temperature. Thermal variation of resistivity depicts non monotonic temperature dependence. The total negative susceptibility increases and band gap of semiconducting $Bi_{0.88}Sb_{0.12}$ samples decreases with increasing temperature.

---

a) Author to whom correspondence should be addressed:
   Electronic mail: arbphy@caluniv.ac.in, Tel: +91-33-23508386 Extn.:429, Fax: +91-33-23519755

## I. Introduction

Semiconducting Bi-Sb alloy is a well known n type thermoelectric material. Bi, having rhombohedral crystal structure, is a semimetal, with overlap energy of 184 meV between the valance and conduction band [1,2]. Bi, however when alloyed with semimetal Sb constitutes a continuous solid solution for the entire range of Sb concentration [3-5]. The Bi-Sb alloy have received special attention due to their attractive physical properties [3, 6-9], emerging from peculiarities of its band structure. Electron-electron and electron-phonon scattering plays an important role in transport properties of this alloy and thus influence its thermoelectric properties [10]. $Bi_{1-x}Sb_x$ alloy shows semiconducting behavior for a small range of Sb content [$(0.07 \leq x \leq 0.22)$]. The band gap ($E_g$) of semiconducting $Bi_{1-x}Sb_x$ alloy depends extensively on Sb concentration and shows a transition from direct ($0.08 \leq x \leq 0.12$) to indirect ($0.12 < x \leq 0.22$) band gap at $x = 0.12$. These are narrow $E_g$ semiconductors with maximum $E_g$ of about 20 meV [3,11]. However, it should be pointed out that several groups have reported maximum $E_g$ around x=0.15 [12,13,14]. But there are different reports also, : *viz.*, Rodionov *et al.* [15] reported maximum $E_g$ around x=0.13, whereas Jain [1] and Malik *et al.* [11] reported for $Bi_{0.88}Sb_{0.12}$ sample. Bi possesses interesting diamagnetic property and the anomalous diamagnetism of elemental Bi has been discussed qualitatively by Adams [16]. The relation between $E_g$ and diamagnetic susceptibility in $Bi_{1-x}Sb_x$ alloy has been elaborately discussed by N. B. Brandt *et al.* [17] M. Sengupta *et al.* studied the effect of inhomogenity on the band parameters and magnetic property of single crystalline semiconducting and semimetallic Bi-Sb alloys [18]. Though limited works have been reported, but studies on magnetic property of the semiconducting $Bi_{1-x}Sb_x$ are still illusive in literature.

Moreover, synthesis and characterisation of semiconducting Bi-Sb polycrystalline alloy is also important from the application point of view. But it is difficult to prepare a homogeneous Bi-Sb alloy. The underlying reason is inherent in the phase diagram [3]. Large temperature difference (ΔT) between the liquidus and the solidus lines, with constitutional super cooling causes macro-segregation of elemental Bi and Sb during synthesis, resulting in microscopic in-homogeneity of the alloy [19,20]. Improvement in homogenization can be achieved by rapidly freezing the homogeneous melt [21] or long time annealing of a quenched alloy [22]. Some methods have been developed to obtain homogeneous single or polycrystalline alloy such as travelling heater method [20], mechanical alloying (MA) [23,24], spark plasma sintering of the MA sample [21,25] etc. It was also reported that, sample synthesized by MA like ball milling technique is bound to increase thermodynamic defects, which in turn might be helpful in increasing the efficiency of a thermoelectric material [24].

In this work, an attempt has been made to study the effect of quenching from different temperatures on $Bi_{0.88}Sb_{0.12}$ alloy. The effect of synthesis temperature on $E_g$ of $Bi_{0.88}Sb_{0.12}$ alloy has been studied. A in depth analysis of the structural properties as well as micro-structural characteristics of the melt quenched alloys are reported here. Finally, the effect of in-homogeneity on the structural, resistive and magnetic property of polycrystalline $Bi_{0.88}Sb_{0.12}$ is also analyzed.

## II. Experiment

Polycrystalline $Bi_{0.88}Sb_{0.12}$ samples were synthesized by melting of high purity Bi and Sb powder. It is an emerging and popular technique for the synthesis of Bi-Sb alloy as well as other related metal chalcogenides based thermoelectric materials [26-29]. Stoichiometric quatities of Bi and Sb powders (each of purity 99.999%; Alfa Aesar, UK) were loaded into an evacuated ($10^{-3}$ torr) quartz tube of 10 mm and 11 mm, inner and outer diameters, respectively. The evacuated quartz tube were heated upto three different temperatures viz., 973 K, 1273 K and 1473 K and soaked for 10 h and finally rapidly quenched in liquid Nitrogen. Powder X-Ray Diffractometer [Model: *X'Pert PRO* (PANalytical)] was used for structural characterization of the synthesized samples. All the X-Ray Diffraction (XRD) measurements were performed in θ-2θ geometry with Cu-K$_α$ radiation. In depth structural analysis was performed using Rietveld (MAUD software) refinement technique [30]. Standard Si was used to determine the instrumental profile [31]. Field Emission Gun Scanning Electron Microscope (FEGSEM: FEI make Helios Nanolab 600i) equipped with a Ocatne Pro Energy Dispersive X-Ray (EDX) detector was used for microstructural and microchemical analysis of the samples. Microhardness measurements were carried out using a Leitz Miniload 2 microhardness tester with an applied load of 100 g and dwell time of 15 seconds. Each reported hardness value is an average of three measurements. Resistivity of all the synthesized samples, as a function of temperature down to 100 K, was measured using conventional four probe method. Silver paste cured at room temperature was used for achieving good ohmic electrical contacts. Differential thermal analysis (DTA) of the synthesized samples was performed in nitrogen atmosphere. Temperature dependent diamagnetic susceptibility of all the melt quenched polycrystalline $Bi_{0.88}Sb_{0.12}$ alloys was measured in the temperature range 30-300K using Superconducting Quantum Interference Device Vibrating Sample Magnetometer (SQUID VSM, Quantum Design).

**III. Result and Discussion**

XRD patterns of the three $Bi_{0.88}Sb_{0.12}$ samples treated at three different temperatures are shown in Figure 1. Variation of the most intense diffraction peak, i.e. (012), for samples quenched from different temperatures, is demonstrated in Figure 1(inset). All the (012) peaks of the synthesized $Bi_{0.88}Sb_{0.12}$ alloys shift to lower angle with increasing temperature. The synthesized samples are soaked at 973 K, 1273 K and 1473 K, respectively followed by rapid quenching at liquid nitrogen. This help the melt quenched samples to retain its phase as obtained at respective temperature. Since treatment temperature is consecutively higher in the three respective samples, the unit cell volume and lattice parameter of the melt quenched samples increases with increasing temperature due to thermal expansion [32]. It is correctly reflected in the shifting of XRD peaks to lower angle. Bi and Sb are isoelectronic and both belong to rhombohedral crystallographic structure with $R\bar{3}m$ space group. In-depth structural analysis has been carried out by Rietveld refinement technique using MAUD software. The experimental XRD pattern and theoretical XRD curve, as obtained after Rietveld refinement, for all the synthesized $Bi_{0.88}Sb_{0.12}$ alloys are presented in Figure S1 (see the supplementary information for Figure S1) [33]. The refinement has been carried out using both atomic position and substitution. Space group $R\bar{3}m$ and point group $D_{3d}$ with hexagonal coordinate system are used for refinement. Refined hexagonal unit cell volume as well as the corresponding lattice parameters, $a_H$ and $c_H$ (within error limit) of the melt quenched samples increases with increasing temperature (Table-1). This is in confirmation with the shift of the XRD peak towards lower angle as observed in Figure 1(inset). The effect of temperature on the strain of the samples is also analyzed. As depicted in table 1, strain increases monotonically with increasing temperature. In addition to Rietveld technique using the MAUD software, Williamson-Hall (WH) plot has also been employed to extract the strain in the synthesized samples. WH analysis, as depicted in Figure S2 (see the supplementary information for Figure S2), indicates similar trend [34]. The saturation vapour pressure of Bi and Sb at around 973 K, 1273 K and 1473 K, are 1 and 10 Pa, 100 and $10^3$ Pa and $10^3$ and $10^4$ Pa, respectively [35]. It is further observed that, the difference between saturation vapour pressures of Sb and Bi increase with increasing temperature. Hence Sb is more volatile than Bi and volatilization of Sb increases with temperature. In the melt quenched samples, Sb is thus segregated in the matrix of Bi-Sb alloy and Sb precipitation increases with increasing temperature. Here, it is noteworthy to mention that Lenoir *et al.* predicted Bi segregation in Bi-Sb alloy and in order to avoid precipitation of Bi in their samples, they adopted travelling heater method as well as mechanical alloying route to synthesise homogeneous Bi-Sb samples [20,23]. On the other hand Robert *et al.* [26] synthesized the samples by quenching at 268 K from 723 K and observed segregation of both Bi and Sb in their Bi-Sb samples. There is a large temperature difference between the liquidus and solidus lines in the phase diagram of Bi-Sb alloy and because of very low diffusion rate in the solid the resulting ingot may exhibit severe segregation [3]. However, the samples reported in the present paper are synthesized by rapid liquid $N_2$ quenching from much higher temperatures (973 K, 1273 K and 1473 K), which might lead to Sb segregation in the Bi-Sb matrix. Precipitation of Sb in Bi-Sb matrix is further evidenced from EDX mapping analysis (discussed later). It is thus quite justified to assume that, the segregation of Sb from the Bi-Sb rhombohedral unit cell leads to the observed strain in the high temperature quenched samples.

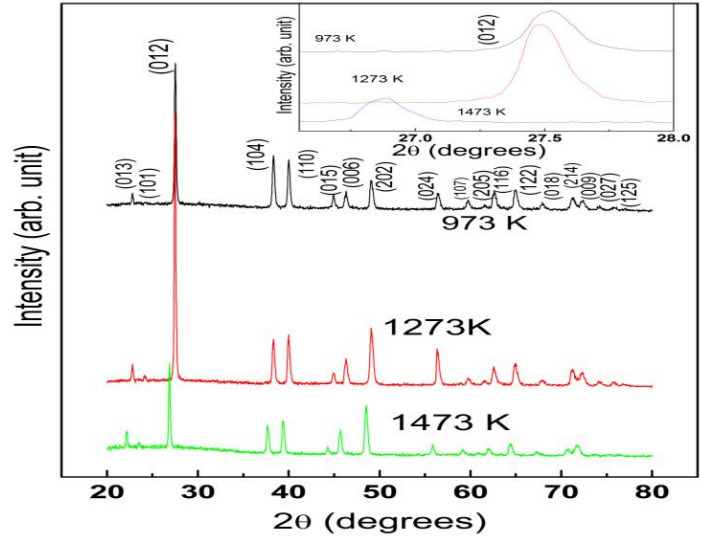

**Fig. 1.** XRD Patterns of $Bi_{0.88}Sb_{0.12}$ alloys quenched from 973 K, 1273 K and 1473 K, respectively. Inset shows (012) peak of rhombohedral $Bi_{0.88}Sb_{0.12}$ alloys.

Rietveld refinement data also indicates that Debye Waller factor [$B_{iso}$], linked to the atomic displacement parameters, increases with temperature (Table-1). Debye-waller factor [$B_{iso}$] consists of two parts, thermal ($B_{thermal}$) and static ($B_{static}$). $B_{static}$ can be calculated using the relation [36]:

$$B_{statics} = 8\pi^2 \langle u_{static}^2 \rangle \quad (1)$$

$$\text{with } \langle u_{static}^2 \rangle = (r_A - r_B)^2 x(1-x) \quad (2)$$

, where $r_A$ and $r_B$ are atomic radii of Bi (0.160 nm) and Sb (0.145 nm), respectively and Sb concentration is represented by 'x'. However, for all the Bi-Sb samples ($Bi_{0.88}Sb_{0.12}$) reported here $r_A$, $r_B$ and 'x' are same, hence $B_{statics}$ is almost unaffected. Thus the variation in $B_{iso}$ reported in Table-1, as obtained from Rietveld refinement, is solely due to $B_{thermal}$, designated as the mean square displacement of atoms from its original position. The large difference in atomic radii of Bi and Sb along with high temperature quenching may cause the off centring of Bi/Sb atoms [37]. Mean square displacement, intimately related to this off centring of Bi/Sb atoms, is thus different in the reported $Bi_{0.88}Sb_{0.12}$ samples and increases subsequently with increasing temperature.

In order to confirm and quantify the segregation of Sb in Bi-Sb matrix, detailed microstructural characterization and EDX mapping of all the synthesized samples are carried out. EDX mapping (Figure 2) clearly indicates segregation of Sb in the Bi-Sb matrix. In order to quantify the amount of Sb segregation, area fraction calculation using *ImageJ*® software is performed. The results have been enumerated in Table-2, indicating that Sb segregation increases with increasing temperature. In addition, an attempt to estimate the grain size of the samples from FEGSEM micrographs is also made. However, it may be indicated here that the sample preparation for microstructural analysis is not trivial. Hence, as an alternative approach hardness measurement is carried out to estimate the effect of temperature on the grain size of the samples. According to Hall-Petch relation, measured hardness of a sample is inversely proportional to the square root of the grain size [38]. These measurements (Table-2) however, do not reveal any significant change in hardness, which in turn implies that, the grain size of the samples do not significantly vary with temperature. In addition, WH method is also employed to estimate the grain size of the samples. WH analysis, as represented in Figure S2 (see the supplementary information for Figure S2) also indicates that grain size do not change significantly with temperature [34]. The results may be understood from the Bi-Sb alloy phase diagram. This diagram indicates that the melting temperature of $Bi_{0.88}Sb_{0.12}$ is 565 K [3,20]. The samples reported in this paper have been quenched from 973 K, 1273 K and 1473 K, all of which are higher than the melting point of $Bi_{0.88}Sb_{0.12}$ alloy. Thus effectively all the samples are quenched from liquid state, which is correctly reflected in the microhardness measurement as well as WH plot.

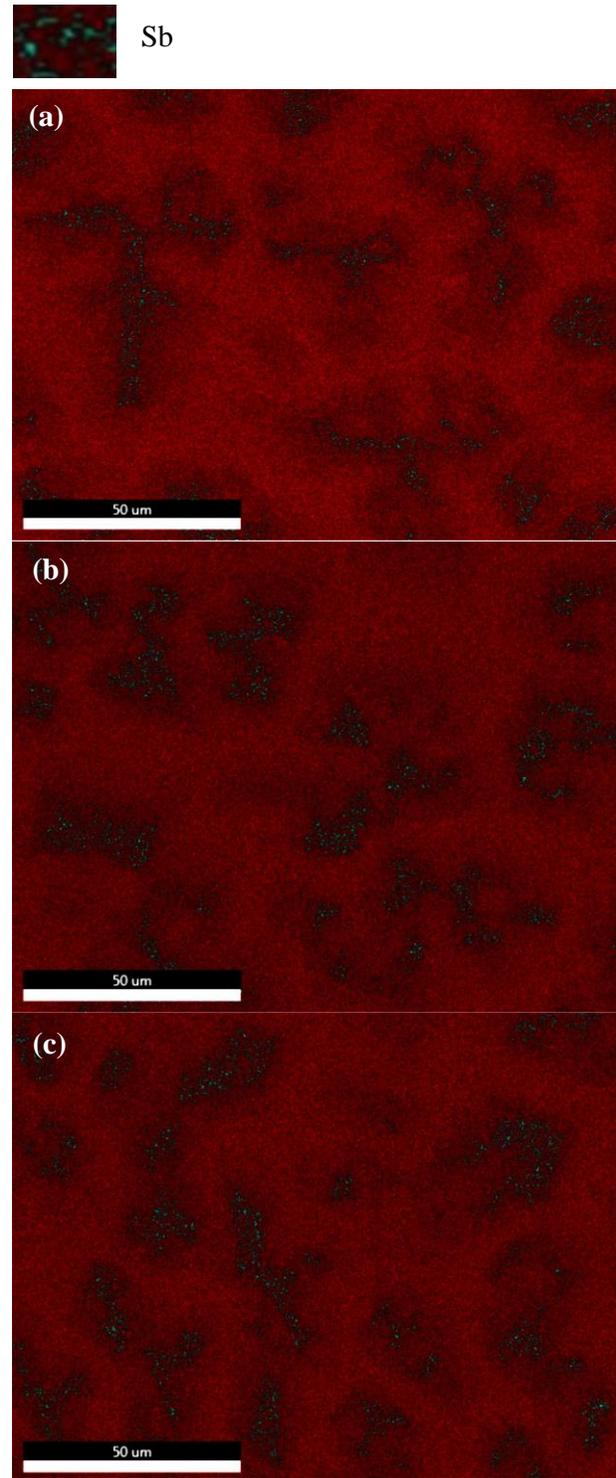

**Fig. 2.** EDX mapping micrograph of $Bi_{0.88}Sb_{0.12}$ samples melt quenched from (a) 973 K, (b) 1273 K, and (c) 1473K.

DTA curves of synthesized samples are given in Figure 3. Melting points of Bi and Sb are 544 K and 903 K, respectively and that of Bi rich $Bi_{0.88}Sb_{0.12}$ is around 565 K [3,20,39]. Thus the endothermic peak observed around 565 K indicates the melting of Bi-Sb alloy. In addition, there is a small endothermic peak at around 885 K for the 1273 K quenched sample. It is due to the melting of Sb and indicates the presence of minute amounts of Sb within the quenched alloy. The strong exothermic peak appearing around 835 K observed for the sample treated at 1473 K is probably due to the formation of some Sb rich phase. Evidence of similar exothermic peak due to the formation of Sb rich $(Bi_{0.25}Sb_{0.75})_2Te_3$ phase has also been reported earlier [39]. Sb segregation increases for the sample treated at 1473 K and form minute amount of Sb rich phase, which is manifested by the exothermic peak around 835 K [Figure 3]. The signature of Sb precipitation in Bi-Sb matrix also corroborates with the EDX mapping analysis [Figure 2, Table 2].

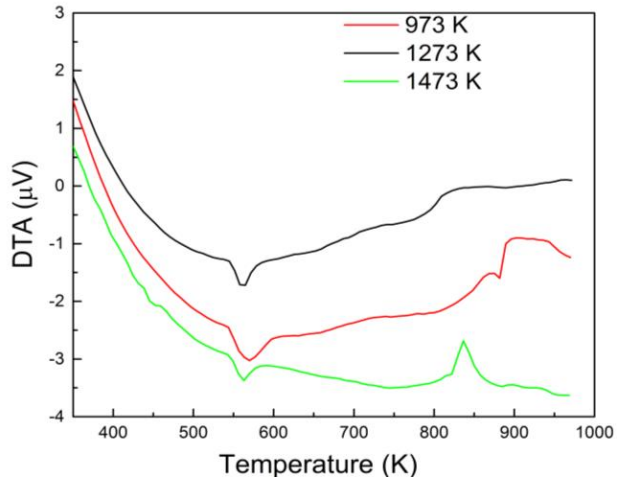

**Fig. 3.** DTA curve for $Bi_{0.88}Sb_{0.12}$ alloys heated at 973 K, 1273 K and 1473 K.

Temperature dependent $\rho$ of $Bi_{0.88}Sb_{0.12}$ alloys, quenched from three different temperatures are presented in Figure 4. The $\rho$-T curves follow similar trend as reported earlier for 973 K air cooled sample, showing non-monotonic temperature dependence with maximum value of $\rho$ at temperature $T_p$ [11]. It is further observed that $\rho$ increases with treatment temperature, which is intimately related to the impurity scattering arising due to Sb

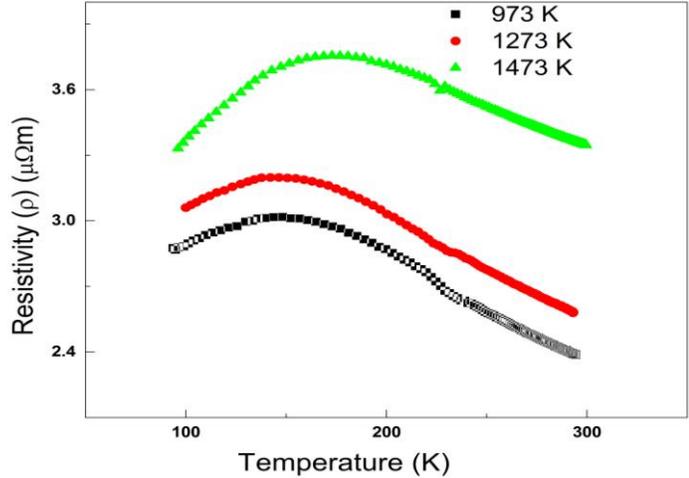

**Fig. 4.** Thermal variation of electrical resistivity ($\rho$) of $Bi_{0.88}Sb_{0.12}$ alloys synthesized by quenching from different

precipitation in the Bi-Sb matrix. The area fraction calculation using *ImageJ®* software clearly revealed that, amount of Sb precipitate increased with temperature. The microscopic inhomogenity arising due to Sb segregation also plays a significant role in the observed increase of $\rho$ with increasing temperature. $E_g$ of the synthesized samples are estimated from the high temperature semiconducting region (for T>$T_p$) of $\rho(T)$ data using the equation[11]

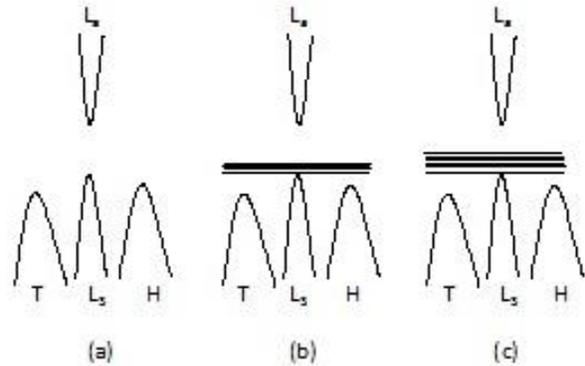

**Fig. 5.** Schematic representation of energy bands near the Fermi level for $Bi_{0.88}Sb_{0.12}$ alloys quenched from (a) 973 K, (b) 1273 K, and (c) 1743 K. For simplicity, the L, T and H point bands are drawn one on top of the others. Evaluation of Sb induced impurity band leads to the decrease of $E_g$ with increasing temperature.

$$\rho = \rho_0 \exp(E_g/2k_B T) \qquad (3)$$

where, $\rho_0$ is constant related to the impurity scattering and $k_B$, the Boltzman constant. The estimated values of $\rho_0$ and $E_g$, extracted from fitting equation 3 with $\rho(T)$ curve, are $1.60 \times 10^{-6}$, $1.81 \times 10^{-6}$, $2.76 \times 10^{-6}$ $\mu\Omega$m and 20.48, 18.37, 12.72 meV, respectively for the samples heated at 973 K, 1273 K and 1473 K. The increase of $\rho_0$ values with increasing temperature clearly suggest the increase of impurity scattering arising due to Sb segregation, as discussed above. Further, it is observed that, $E_g$ decreases with increasing temperature. It is noteworthy to mention that for pristine $Bi_{0.88}Sb_{0.12}$ alloy, $E_g$ arises due to the gap between $L_a$ and $L_s$ bands [1,6,11]. However, impurity levels due to Sb have been postulated to exist between the $L_a$ and $L_s$ bands [18]. Thus in validation of the observation made from in depth microstructural and microchemical analysis, it can be commented that, increased amount of Sb precipitation in the higher temperature melt quenched samples give rise to impurity levels between $L_a$ and $L_s$ bands. This is agreement with the experimental observations of Brandt *et al.* [17]. With increasing temperature segregation of Sb increases and changes the position of the associated impurity levels such that $E_g$, as estimated from the $\rho(T)$ data, decreases. Schematic representation of the energy band diagram near the Fermi level, showing the evolution of corresponding bands for the $Bi_{0.88}Sb_{0.12}$ samples, is depicted in Figure 5.

In order to investigate the diamagnetic property of the melt quenched $Bi_{0.88}Sb_{0.12}$ samples, temperature dependent magnetic susceptibility ($\chi$) measurement is performed [Figure 6]. All the samples are diamagnetic in nature. For all the melt quenched $Bi_{0.88}Sb_{0.12}$ alloys, $\chi$ increases monotonically with increasing temperature. $\chi(T)$ behavior shows that, initially it increases slowly with increasing temperature, but beyond a certain temperature $T_p^{'}$ ($T_p^{'} \sim 150$ K), it increases at a much faster rate. Interestingly it is found that, $T_p$ as estimated from $\rho(T)$ curve, almost coincides with $T_p^{'}$.

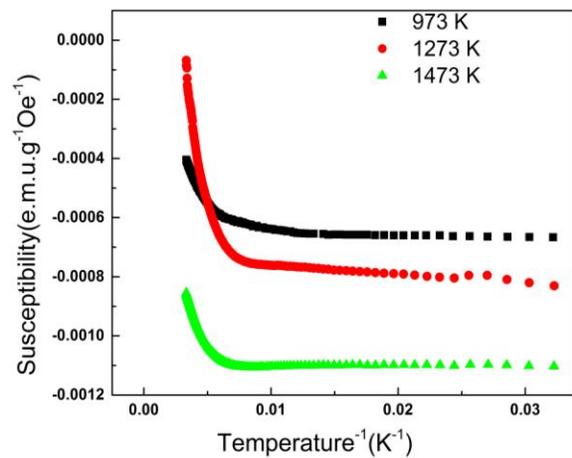

**Fig. 6.** Magnetic susceptibility as a function of 1/T for different melt quenched $Bi_{0.88}Sb_{0.12}$ alloys.

For semiconducting Bi-Sb alloys the total negative $\chi$ mainly arises from the valance band electrons producing strong diamagnetism ($\chi_G$) and the thermally activated carriers, which contribute to a paramagnetic term ($\chi_C$) [40]. However, according to Buot and McClure [41] total $\chi$ in semiconducting Bi and related alloys can be written as:

$$\chi = \chi_A + \chi_C + \chi_G, \qquad (4)$$

where, $\chi_A$ is ionic susceptibility, and $\chi_C$ arises from the carrier present in system (i.e. electrons at L point and holes at H and T point).

It should be noted that, $\chi_C$ strongly depends on the position of Fermi surfaces and the chemical potential ($\mu$) [17]. $\mu$ changes with temperature and is related to the density of states of the conduction and valence bands. For the reported Bi-Sb alloys, $L_s$ and H bands, where, H is the heavy hole band, are almost at the same position. Excitation of electrons from the $L_s$ and H bands to the conduction band ($L_a$) increases the number of holes in valence band ($L_s$ and H band). For $T > T_p$, population of carriers in the system thus increases [10], which further increases $\chi_C$. In fact $\chi(T)$ data for individual samples indicates the increase of susceptibility ($\chi$) at a faster rate for $T > T_p^{'}$, signifying that the magnitude of overall susceptibility decreases due to the increased positive paramagnetic contribution from the conduction electrons. Figure 6 further depicts that, magnitude of total negative $\chi$, i.e., diamagnetic susceptibility increases with increasing temperature. Qualitatively, diamagnetism of Bi can be represented as $\chi_0/\gamma$, where $\chi_0$ is the susceptibility of a typical metal obtained from Pauli-Landau formula and γ is the thermal gap ($E_g$) [17]. Here it should be recalled that, for the synthesized samples $E_g$ decreases with increasing temperature, which is in corroboration with the observed susceptibility data. Brandt *et al.* also reported similar observations [17].

## IV. Conclusions

In conclusion, the effect of treatment temperature on melt quenched $Bi_{0.88}Sb_{0.12}$ is reported. Polycrystalline $Bi_{0.88}Sb_{0.12}$ alloys are heated at three different temperatures, followed by rapid liquid nitrogen quenching. Lattice parameter, lattice strain and $B_{iso}$ of the synthesized samples increase with increasing temperature. Minute amounts of Sb segregation from Bi-Sb unit cell in the samples quenched from different temperatures might have led to the observed lattice strain. The Sb precipitation, which increases with increasing temperature, is also evidenced from DTA measurement and EDX mapping. The magnitude of diamagnetic susceptibility increases with increasing temperature. The ρ(T) data, showing non-monotonic temperature dependence, also increases with increasing sample treatment temperature and is related to the increased impurity scattering as well as microscopic inhomogeneity arising due to Sb precipitation. On the other hand $E_g$, estimated from the high temperature semiconducting regime, decreases with increasing temperature. It is thus postulated that, the impurity levels due to Sb lie between $L_a$ and $L_s$ bands. The Sb segregation increases with increasing temperature and influences the associated impurity levels in such a way that $E_g$ decreases.

**Acknowledgement:**

This work is supported by Department of Science and Technology (DST), Govt. of India; UGC, Govt. of India and UGC-DAE CSR, Kalpakkam node in the form of sanctioning research project, reference no. SR/FTP/PS-25/2009, 39-990/2010(SR) and CSR-KN/CRS-65/2014-15/505, respectively. Author KM and DD is thankful to UGC and UGC-DAE CSR, Govt. of India, respectively for providing Research Fellowships. Authors are also thankful to Prof. A. Ghosh and Dr. Saptarshi Biswas of Department of Chemistry, University of Calcutta for the DTA measurement.

**Table-1**. Structural parameters of hexagonal unit cell of the $Bi_{0.88}Sb_{0.12}$ alloy melt quenched from 973 K, 1273 K and 1473 K respectively, as obtained by Rietveld Refinement technique using MAUD software along with the corresponding $B_{iso}$, strain and Goodness of fit (GoF) or $\chi^2$ value.

| Temperature (K) | Hexagonal lattice parameters | | Volume ($\text{Å}^3$) | $B_{iso}$ | Strain | GoF or $\chi^2$ |
|---|---|---|---|---|---|---|
| | $a_H$ (Å) | $c_H$ (Å) | | | | |
| 973 | 4.5340 ± 3.0 x $10^{-4}$ | 11.8380 ± 2.0 x $10^{-3}$ | 210.7533 ± 40.7 x $10^{-3}$ | 0.469 | 1.90 x $10^{-3}$ ± 4.9 x $10^{-5}$ | 1.186 |
| 1273 | 4.5351 ± 5.4 x $10^{-4}$ | 11.8361 ± 1.8 x $10^{-3}$ | 210.82169 ± 48.2 x $10^{-3}$ | 1.420 | 1.94 x $10^{-3}$ ± 5.5 x $10^{-5}$ | 1.377 |
| 1473 | 4.5368 ± 3.2 x $10^{-4}$ | 11.8410 ± 1.2 x $10^{-3}$ | 211.06712 ± 30.5 x $10^{-3}$ | 1.520 | 2.10 x $10^{-3}$ ± 3.4 x $10^{-4}$ | 1.094 |

**Table-2.** The results of WH analysis, hardness measurement and area fraction of precipitated Sb (using *ImageJ* ® software) as obtained for $Bi_{0.88}Sb_{0.12}$ samples quenched from different temperatures.

| Temperature (K) | WH Result | | Microhardness (VHN) | Area faction (%) of Sb |
|---|---|---|---|---|
| | Strain | Grain size (nm) | | |
| 973 | 4.55 x $10^{-3}$ | 426 ± 32 | 33.6 | 1.52 |
| 1273 | 4.63 x $10^{-3}$ | 412 ± 26 | 31.7 | 2.55 |
| 1473 | 4.74 x $10^{-3}$ | 441 ± 36 | 33.0 | 5.28 |